\documentclass[11pt]{article}
\usepackage{amsmath,amssymb,color,graphics,epsfig,cite}

\usepackage{amsfonts}

\makeatletter
\@addtoreset{equation}{section}
\makeatother

\newcommand{\be}{\begin{equation}}
\newcommand{\ee}{\end{equation}}
\newcommand{\bea}{\setlength\arraycolsep{2pt} \begin{eqnarray}}
\newcommand{\eea}{\end{eqnarray}}
\newcommand{\nn}{\nonumber}

\def\fft#1#2{{\frac{#1}{#2}}}

\def\0{{\sst{(0)}}}
\def\1{{\sst{(1)}}}
\def\2{{\sst{(2)}}}
\def\3{{\sst{(3)}}}
\def\4{{\sst{(4)}}}
\def\5{{\sst{(5)}}}
\def\6{{\sst{(6)}}}
\def\7{{\sst{(7)}}}
\def\8{{\sst{(8)}}}
\def\sst#1{{\scriptscriptstyle #1}}

\begin{document}

\begin{center}
{\Large {\bf Classical Holographic Relations and Alternative Boundary Conditions for Lovelock Gravity}}

\vspace{20pt}

{\large H. Khodabakhshi and H. L\"u}

\vspace{10pt}

{\it Center for Joint Quantum Studies and Department of Physics\\ School of Science, Tianjin University,\\ Yaguan Road 135, Jinnan District, Tianjin 300350, China}

\vspace{40pt}

\underline{ABSTRACT}
\end{center}

We obtain the classical holographic relation for the general Lovelock gravity and decompose the full Lagrangian into the bulk term and the surface term, expressed as a total derivative $\partial_\mu J^\mu$. By classical holographic relation, we mean that $J^\mu$ is determined completely by the bulk term. We find that the bulk term is not degenerate, or first-order in this foliation-independent approach. We then consider the Arnowitt-Deser-Misner (ADM) formalism where the foliation coordinate $w$ is treated as special. We obtain the classical holographic-degenerate relation with the first-order bulk term that does not involve higher than one derivative of $w$.  For Einstein gravity, the two approaches lead to the same bulk term, but different ones for higher-order Lovelock gravities. The classical  holographic-degenerate formulation in the ADM approach allows us to consider alternative boundary conditions in the variation principle with different Myers terms.  We show in the semiclassical approximation that the black hole entropy in all cases is the same as the one obtained under the standard Dirichlet boundary condition. We also generalize the formalism to general $f(L_{\rm Lovelock}^{(k)})$-gravity.

\vfill{\footnotesize  h.khodabakhshi@ipm.ir \ \ \ mrhonglu@gmail.com}


\thispagestyle{empty}
\pagebreak

\tableofcontents
\addtocontents{toc}{\protect\setcounter{tocdepth}{2}}

\section{Introduction}

	Holographic principle, relating theories in $D$ and $D-1$ dimensions \cite{Susskind:1994vu, tHooft:1999rgb}, has played a significant role in theoretical physics since the discovery of the AdS/CFT correspondence \cite{Maldacena:1997re}, a conjectured duality based on string theory. The  holographic principle is quantum in nature and in a suitable limit can relate a classical gravity theory to its strongly coupled dual quantum theory. In this paper we would like to explore a very different type of holographic relation at the level of Lagrangian \cite{pad1, pad2} and we would like to call it ``classical holographic relation'', to distinguish it from the usual ``holographic duality'' or ``holography''. We can decompose a higher-derivative Lagrangian into a bulk and surface term as
	\begin{equation}
		\mathcal{L} (\Phi, \partial \Phi, \partial^2 \Phi,...)=\mathcal{L}_{\rm {bulk}} (\Phi, \partial \Phi, \partial^2 \Phi,..)+\partial_{\mu} J^{\mu} (\Phi, \partial \Phi, \partial^2 \Phi,..),
		\label{hhh}
	\end{equation}
where $\Phi$'s are dynamical fields. By holography we mean $J^{\mu}$ is not arbitrary but it is completely determined by the bulk Lagrangian. In general relativity (GR), since the affine connection $\Gamma$, which is a derivative of the metric tensor, is not tensorial under the general transformation, the Einstein-Hilbert (EH) Lagrangian involves not only $\Gamma$ but also its higher derivative $\partial \Gamma$. It turns out, miraculously, that the EH Lagrangian can be written in the form of (\ref{hhh}) with
	\begin{equation}
		\mathcal{L}_{\rm {bulk}}=\mathcal{L}_{\rm quad} (\Gamma)=\mathcal{L}_{\rm quad} (g, \partial g)
		\label{hhh1}
	\end{equation}
{\it i.e.}~it is quadratic in $\Gamma$ or $ \partial g$ only \cite{pad1,pad2,HKH1}, analogous to classical mechanics or field theories. In other words, although the EH Lagrangian involves higher-derivatives on the metric, they can all be absorbed into a total derivative that is fully determined by the first-order bulk Lagrangian. In classical mechanics according to the Ostrogradsky theorem \cite{Os}, a Lagrangian is degenerate if one can write it as a sum of a first-order Lagrangian and a total derivative term. Hence if the bulk term of the classical holographic relation was a first-order Lagrangian, we would like to call it classical holographic-degenerate. Clearly, in Einstein gravity, the classical dynamics are determined by the bulk Lagrangian and the total derivative term plays the role of making the action invariant under the general transformation. Thus the classical  holographic relation and general diffeomorphism are closely related.

When a Lagrangian contains two or more derivatives on a dynamical field, it becomes subtle to obtain the equation of motion {\it via} the variation principle. Naively, one would have to impose not only the Dirichlet boundary condition (BC) but also the Neumann BC, where the momentum fields vanish on the boundary. However, it is inconsistent to allow a canonical pair with non-vanishing Poisson bracket to vanish simultaneously, even only on the boundary \cite{HKH1,reggi,Hint}. In Einstein gravity, this issue was resolved by introducing appropriate Gibbons-Hawking-York (GHY) surface term \cite{Gib,York}, whose coefficient is independent of the space-time dimensions of the theory. The variation principle then becomes consistent by simply imposing the Dirichlet BC. However, the classical holographic degeneracy allows one to impose alternative BCs, since for degenerate Lagrangian, we can define the momentum fields straightforwardly from the first-order bulk Lagrangian. It turns out that one can consistently impose the Neumann BC in four dimensions without having to introduce any GHY term. In general dimensions, an appropriate GHY term with a specific dimension-dependent coefficient is needed for the Neumann BC \cite{Krishnan, HKH1}.

In this paper, we would like to explore further the classical holographic relation in higher-derivative gravities that are constructed from general Riemann tensor invariants. We would like to focus on degenerate theories which necessarily require ghostfree combinations of Riemann tensor invariants. One such a theory is $f(R)$-gravity. However, it was shown that the theory is not degenerate, but can be made so using the Ostrogradsky approach \cite{Os}. Specifically, one can follow the approach and introduce a scalar field in the framework of Brans-Dicke formalism \cite{Cap,Sot,Far}. In this different but equivalent formalism, the Lagrangian becomes degenerate and allows to have different BCs with appropriate GHY terms \cite{HKH1}.

The more interesting ghostfree higher-derivative gravity is perhaps Lovelock gravity, which is a specific higher-order polynomial combinations of the Riemann tensor such that the field equation remains two derivatives. The Lovelock series can be classified by the homogeneous polynomial order $k$ of the Riemman tensor, with $k=1,2$ corresponding to the familiar Einstein and Gauss-Bonnet (GB) gravities respectively. Lovelock gravity is also a function of $\Gamma$ and $\partial \Gamma$.

We find that general Lovelock gravity has the classical holographic relation as in (\ref{hhh}), but the  bulk Lagrangian is not degenerate, in that it is a function not only of $\Gamma$ but also of $\partial\Gamma$. One difference between the bulk term and the full Lovelock Lagrangian is that the highest polynomial order of $\partial \Gamma$ in the bulk is one less. Consequently, the bulk Lagrangian of Einstein gravity with $k=1$ does not have $\partial\Gamma$. Using the foliation-independent approach, the classical holographic relation of Lovelock gravities was also studied in \cite{pad1,pad2}, but our results are different.

Thus the  foliation-independent approach where the surface term is $\partial_\mu J^\mu$ leads to non-degenerate bulk Lagrangian for general Lovelock gravities.  This leads us to consider a different approach based on the ADM formalism \cite{ADM1, ADM2}. We can choose special coordinates $x^\mu=(w, x^a)$ such that $J^\mu = (J^w,0)$. We then decompose the metric using the ADM formalism, treating the coordinate $w$ as the foliation coordinate.  We then find that Lovelock gravity is not only classical holographic but also degenerate in the sense that the bulk Lagrangian has no more than one derivative of the coordinate $w$. The two-derivative terms in the bulk Lagrangian are all associated with coordinates $x^a$, entering through the Riemann tensor of the sub-manifold. We express the surface term as $\partial_w {\cal L}_{\rm surf}$. It turns out that the bulk Lagrangian is precisely the first-order Lagrangian obtained using the ADM formalism by Teitelboim and Zanelli \cite{Tel} and recently considered in a straightforward manner in \cite{Pap,Fel}. Our result is to establish the classical holographic relation. For Einstein gravity $(k=1)$, we find that the bulk Lagrangian is identically the same for both the foliation-independent and the ADM approaches. They are however not the same for GB ($k=2$) or higher-$k$ Lovelock gravities. The classical holographic degeneracy in the ADM formalism allows us to reexamine the BCs in the variation principle for general Lovelock gravities.
	
The paper is organized as follows. In section \ref{2}, we obtain the classical holographic-degenerate relation for the Lovelock Lagrangian by the ADM formalism, making use of the Teitelboim-Zanelli Lagrangian which is the first-order Lagrangian \cite{Tel,Pap} and Myers boundary term \cite{Mayers}. From the classical holographic-degenerate relation, we add appropriate Myers terms to the Lovelock action to make the variation principle well defined under Dirichlet or Neumann BCs. We then consider $f(L_{\rm Lovelock}^{(k)})$-gravity and show that it has no classical holographic relation. Using Ostrogradsky approach in the frame work of Brans-Dicke formalism, we find that its equivalent scalar-tensor theory can be written in a degenerate form. This enables us to introduce the consistent {Myers} terms for Dirichlet, Neumann and two types of mixed BCs in arbitrary dimensions. It is intriguing to observe that there exists a mixed BC that does not require any Myers term in all dimensions. Furthermore GHY and Myers terms play an important role in the calculation of the black hole entropy \cite{HKH2, Osh, Rob, Pad}. Using semiclassical approximation method \cite{Hint, Brown, HKH2}, we examine what happens to the black hole entropy under different BCs as well as different Myers terms. We expect physically the entropy should be the same under different BCs and our results indeed confirm this. In section 3, we investigate classical holographic relation for the Lovelock Lagrangian based on the foliation-independent approach \eqref{hhh}. We give explicit bulk Lagrangian and the classical holographic relation.  We consider the Friedmann-Lema\v itre-Robertson-Walker (FLRW) cosmological metric to illustrate the difference between the ADM formalism and the foliation-independent approach.

\section{Classical holographic Lagrangian in ADM formalism}

\label{2}
We begin with the general Lagrangian of the Lovelock series, given by \cite{Love}
\be
{\cal L}=\sqrt{-g} \mathit{L}_{\rm Lovelock}= \sqrt{-g} \sum_{k\ge 0} \alpha_{(k)}\, \mathit{L}_{\rm Lovelock}^{(k)}\,,\label{lovelock}
\ee
where $\alpha_{(k)}$ are dimensionful coupling constants and $\mathit{L}_{\rm Lovelock}^{(k)}$ is the $k$'th order Lovelock combinations of the Riemann tensor, given by
\be
\mathit{L}_{\rm Lovelock}^{(k)} = \fft{(2k)!}{2^k}\,
\delta^{\mu_1\cdots \mu_{2k}}_{\nu_1\cdots \nu_{2k}}\,
R^{\nu_1\nu_2}_{\mu_1\mu_2}\, R^{\nu_3\nu_4}_{\mu_3\mu_4} \cdots
R^{\nu_{2k-1}\, \nu_{2k}}_{\mu_{2k-1}\, \mu_{2k}}\,.\label{Ekdef}
\ee
The multi-index Kronecker delta symbol is defined to be totally antisymmetric, {\it i.e.}
\be
\delta^{\mu_1\cdots \mu_{2k}}_{\nu_1\cdots \nu_{2k}} =
\delta^{[\mu_1} _{\nu_1}\, \delta^{\mu_2}_{\nu_2}\cdots
\delta^{\mu_{2k}]}_{\nu_{2k}}=\delta^{[\mu_1} _{[\nu_1}\, \delta^{\mu_2}_{\nu_2}\cdots
\delta^{\mu_{2k}]}_{\nu_{2k}]}\,,
\ee
where the square brackets denote conventional unit-strength antisymmetrisations (so, for example, $X^{[\mu_1\cdots \mu_p]} = X^{[[\mu_1\cdots \mu_p]]}$).  Note
that with our choice of normalization, we have
\be
L_{\rm Lovelock}^{(k)} = R^k +\cdots \,,
\ee
with unit coefficient for the purely Ricci scalar term, where the
ellipses denote all terms involving one or more
uncontracted Ricci tensor or Riemann tensor.  Therefore, we have $L^{(0)}=1$,
$L^{(1)}=R$ and $L^{(2)}= L_{\rm GB}\equiv R^2 - 4 R^{\mu_1 \nu} R_{\mu \nu} + R^{\mu \nu \rho \sigma} \, R_{\mu \nu \rho \sigma}$, {\it etc}.
The equations of motion is $\sum_k E^{(k)}_{\mu\nu}=0$, with
\be
E^{\mu(k)}_\nu=-\frac{(2k)!}{2^{k+1}} \delta^{\mu \mu_1\dots \mu_{2k}}_{\nu \nu_1 \dots \nu_{2k}} R^{\nu_1 \nu_2}_{\mu_1 \mu_2} \dots R^{\nu_{2k-1} \nu_{2k}}_{\mu_{2k-1} \mu_{2k}}.
\label{eom}
\ee

\subsection{The classical holographic relation}

In this section, we would like derive a classical holographic relation of general Lovelock gravity in the ADM formalism, where the metric is decomposed as
\begin{equation}
	ds^2=\epsilon N(\omega,x^a)^2 dt^2+h_{ab}(w, x^a) \big(dx^a + N^a(\omega,x^a) dt\big)\big(dx^b + N^b(\omega,x^a) dt\big),
	\label{metr}
\end{equation}
in which the coordinate $w$ is specially treated, as a foliation coordinate. (The foliation-independent approach will be given in section 3.) For constant $w$ on the $\partial \mathcal{M}$ of $x^a$, the sub-manifold is spacelike when $\epsilon=-1$ and timelike when $\epsilon=+1$. This metric can be further simplified to the Gaussian metric under appropriate gauge fixing and the metric has the following form \cite{Guas}
\be
ds^2=\epsilon N(\omega)^2d\omega^2 +h_{ab} (\omega,x^a) dx^a dx^b.
\label{s02}
\ee
We focus on our discussion using the Gaussian metric.

It is clear that all the metric components enter the Lovelock Lagrangian with either no $w$ derivative, one derivative $\partial_w$ or two derivatives $\partial^2_w$. We would like first to split the Lagrangian into two parts: terms that involve no $\partial^2_w$ and terms that contain $\partial_w^2$. The key to the classical holographic relation is that the latter part of the Lagrangian is actually a total derivative, {\it i.e.}~it is a surface term. In other words, we need to establish that the full Lagrangian can be expressed as
\be
\sqrt{-g} L^{(k)}= N\sqrt{|h|} L^{(k)}_{{\text{ADM}}}+\partial_\omega \mathcal{L}_{\rm surf}^{(k)} \label{dec}\,.
\ee
The first term was obtained in \cite{Tel}, given by
\be
\sqrt{-g} L^{(k)}_{{\text{ADM}}}=N\sqrt{|h|}  \sum_{l=0}^{k} C_{l}^{(k)} \delta^{{a_1\dots a_{2k}}}_{b_1\dots b_{2k}} R^{b_1 b_2}_{a_1 a_2}\dots R^{b_{2l-1} b_{2l}}_{a_{2l-1} a_{2l}} K^{b_{2 l+1}}_{b_{2 l+1}} \dots K^{b_{2k}}_{b_{2k}},
\label{lquad}
\ee
in which the coefficient is
\be
C^{(k)}_{l}=\frac{k! (2k)! 2^{k-2l} \epsilon^{k-l}}{l! (2(k-l)-1)!!},
\ee
and $K$ is the extrinsic curvature
\be
K_{ab}=\frac{1}{2N}\partial_\omega h_{ab}.
\label{s03}
\ee
In order to see explicitly that $L^{(k)}_{{\text{ADM}}}$ is absent from having $\partial_w^2$, we can adopt the Gauss-Codazzi-Mainardi identity to express the Riemann tensor as
\be
R^{a_1 a_2}_{b_1 b_2}=\bar{R}^{a_1 a_2}_{b_1 b_2}- \epsilon \big(K^{a_1}_{b_1} K^{a_2}_{b_2}+K^{a_1}_{b_2} K^{a_2}_{b_1}\big),
\label{s0}
\ee
where $\bar R^{ab}_{cd}$ is Riemann tensor on $h_{ab}$.

By classical holographic relation, we mean that the surface Lagrangian ${\cal L}^{(k)}_{\rm surf}$ is specified by the bulk one.  To find this relation, it is instructive to write the two terms as \cite{Pap, Pap1}
\bea
L^{(k)}_{{\text{ADM}}}&=&\frac{(2k)!}{2^k} \delta^{a_1\cdots a_{2k}}_{b_1\cdots b_{2k}}\,
R^{b_1b_2}_{a_1a_2}\, R^{b_3b_4}_{a_3a_4} \cdots
R^{b_{2k-1}\, b_{2k}}_{a_{2k-1}\, a_{2k}} + X\,,\nn\\
\partial_\omega \mathcal{L}_{\rm surf}^{(k)}&=& \frac{ (2k)! k N \sqrt{|h|}}{2^{k-2}}\delta^{a_1\cdots a_{2k-1}}_{b_1\cdots b_{2k-1}}\bigg(R^{\omega b_1}_{\omega b_2}R^{b_2 b_3}_{a_2 a_3} +(k-1) R^{\omega b_1}_{a_1 b_2} R^{b_2 b_3}_{\omega a_3} \bigg)R^{b_4 b_5}_{a_4 a_5} \dots R^{b_{2k-2} b_{2k-1}}_{a_{2k-2} a_{2k-1}}\nn\\
&&-N \sqrt{|h|} X\,.
\eea
where
\bea
X &=&(2k)! 2k \int_0^1 ds\, \delta^{a_1\cdots a_{2k}}_{b_1\cdots b_{2k}} K^{b_1}_{a_1} K^{b_2}_{a_2} \bigg(\frac{1}{2} \bar{R}^{b_3b_4}_{a_3a_4}-s^2 \epsilon K^{b_3}_{a_3} K^{b_4}_{a_4} \bigg) \times \cdots
\nonumber\\
&&\dots \times \bigg(\frac{1}{2} \bar{R}^{b_{2k-1}b_{2k}}_{a_{2k-1}a_{2k}}-s^2 \epsilon K^{b_{2k-1}}_{a_{2k-1}} K^{b_{2k}}_{a_{2k}} \bigg).
\eea
One can show $\partial_\omega \mathcal{L}_{\rm surf}^{(k)}$ is indeed a total derivative by finding explicitly the surface Lagangian $\mathcal{L}_{\rm surf}^{(k)}$. To do so, one uses the first-order Lagrangian, and obtains the associated canonical momentum $\pi ^{ab} _{(k)}$ conjugate to $h_{ab}$ as follows
\bea
\pi ^{ab(k)} &\equiv & \frac{\partial (N\sqrt{|h|} L^{(k)}_{{\text{ADM}}})}{\partial (\partial_\omega h_{ab})}\nn\\
&=& (2k)! k \epsilon \sqrt{|h|}
 \int_0^1 ds\, \delta^{a_1\cdots a_{2k-1} a}_{b_1\cdots b_{2k-1} c} h^{cb} K^{b_1}_{a_1}
\bigg(\frac{1}{2} \bar{R}^{b_2b_3}_{a_2a_3}-s^2 \epsilon K^{b_2}_{a_2} K^{b_3}_{a_3} \bigg) \times\dots
\nonumber\\
&& \dots \times \bigg(\frac{1}{2} \bar{R}^{b_{2k-2}b_{2k-1}}_{a_{2k-2}a_{2k-1}}-s^2 \epsilon K^{b_{2k-2}}_{a_{2k-2}} K^{b_{2k-1}}_{a_{2k-1}} \bigg).
\label{momenta}
\eea
One can then establish that the trace of the canonical momentum has the following form \cite{Tel, Pap, Aim}
\begin{align}
\pi^{(k)}=h_{ab} \pi ^{ab(k)} =-\frac{D-2k}{2} \mathcal{L}_{\rm surf}^{(k)}
\label{moment}
\end{align}
where  ${\cal L}_{\rm surf}^{(k)}$ is the Myers boundary term under Dirichlet BC and it is given by  \cite{Mayers} (see also \cite{Liu:2017kml},)
\begin{align}
\mathcal{L}_{\rm surf}^{(k)}=&- (2k)! 2k \epsilon \sqrt{|h|} \int_0^1 ds\, \delta^{a_1\cdots a_{2k-1}}_{b_1\cdots b_{2k-1}} K^{b_1}_{a_1} \bigg(\frac{1}{2} \bar{R}^{b_2b_3}_{a_2a_3}-s^2 \epsilon K^{b_2}_{a_2} K^{b_3}_{a_3} \bigg)\times \dots
\nonumber\\
& \times \bigg(\frac{1}{2} \bar{R}^{b_{2k-2}b_{2k-1}}_{a_{2k-2}a_{2k-1}}-s^2 \epsilon K^{b_{2k-2}}_{a_{2k-2}} K^{b_{2k-1}}_{a_{2k-1}} \bigg)=
\nonumber\\
&
\sqrt{|h|} \sum_{l=0}^{k-1}\mathcal{W}^{(k)}_{l} \delta^{a_1\cdots a_{2k-2l-2} a_{2k-2l-1}\cdots a_{2k-1}}_{b_1\cdots b_{2k-2l-2} b_{2k-2l-1}\cdots b_{2k-1}}\nonumber \\ &(\bar{R})^{b_1b_2}_{a_1a_2}\, \cdots
(\bar{R})^{b_{2k-2l-3}\, b_{2k-2l-2}}_{a_{2k-2l-3} a_{2k-2l-2}} \hspace{.2cm}K^{b_{2k-2l-1}}_{a_{2k-2l-1}}\cdots K^{b_{2k-1}}_{a_{2k-1}},
\end{align}
where the coefficient is
\be
\mathcal{W}^{(k)}_{l}=\frac{(2k)! (k-1)! \epsilon^{k-l} (-2)^{l}} {2^{k-1} (k-l-1)!l!(2l+1)}.
\ee
Now using Eqs.~(\ref{momenta}), (\ref{moment}) and (\ref{dec}) we obtain
\be
\sqrt{-g} \mathit{L}_{\rm Lovelock}^{(k)}(g,\partial g,\partial^2 g)=N \sqrt{h} L^{(k)}_{{\text{ADM}}}(h,\partial h)+\frac{2}{D-2k} \partial_\omega\bigg(-h_{ab} \frac{\partial (N\sqrt{h} L^{(k)}_{{\text{ADM}}})}{\partial (\partial_\omega h_{ab})}\bigg),
\label{hol}
\ee
which is the classical holographic relation for the general Lovelock Lagrangian in the ADM formalism.  This relation establishes that the total derivative term is specified by the bulk term. Note that when $D=2k$, the second term dominates, indicating that the theory is a total derivative.

\subsection{Variational principle}\label{3}

For the Einstein-Hilbert action or its high-derivative generalization including Lovelock gravity, in order to obtain the equations of motion {\it via} the variation principle, one would appear to have to impose both the Dirichlet and Neumann BCs. However, we are not allowed to set the canonical pair, with non-vanishing Poisson brackets, vanish simultaneously, even only on the boundary. The remedy is to introduce some appropriate surface terms to the action so that the variational principle is well defined \cite{Gib,York}.

As reviewed in the previous subsection, in Gaussian coordinates and generally in the ADM formalism Lovelock theory is described by a degenerate Lagrangian, {\it i.e.}~it can be written as the sum of the  first-order Lagrangian and the total derivative term. Using the classical holographic relation (\ref{hol}) and writing the boundary terms of the action in terms of fields and momentum fields will enable us to introduce the consistent Myers term for the traditional Dirichlet BC or the new alternative Neumann BC.

Considering Eq.(\ref{hol}), and taking variation of the action
\be
\mathcal{A}^{(k)}_{\rm} = \int_{\mathcal{M}} d^Dx\, \sqrt{-g}\, \mathit{L}^{(k)}_{\rm Lovelock}=\int_{\mathcal{M}} d^{D}x \mathcal{L}^{(k)}_{{\text{ADM}}}-\frac{2}{D-2k}\int_{\partial _{\mathcal{M}}}d^{D-1} x h_{ab} \pi^{ab(k)}\,,
\ee
yields
\bea
	\delta\mathcal{A}^{(k)}&=&\int_{\mathcal{M}}d^{D}xE^{\mu\nu(k)}\delta g_{\mu\nu}-\frac{2}{D-2k}\Big(\int_{\Sigma_t} d^{D-1}yh_{ab(\Sigma)}\delta \mathit{\pi}_{(\Sigma)}^{ab(k)}+\int_{\mathcal{B}} d^{D-1}z \,h_{ab(\mathcal{B})}\delta \mathit{\pi}_{(\mathcal{B})}^{ab(k)}\Big)\nonumber\\
	&&
+\frac{D-2(k+1)}{D-2k}\Big(\int_{\Sigma_t}d^{D-1}y\mathit{\pi}_{(\Sigma)}^{ab(k)}\delta h_{ab(\Sigma)}+ \int_{\mathcal{B}}d^{D-1}z\, \mathit{\pi}^{ab(k)}_{(\mathcal{B})}\delta h_{ab(\mathcal{B})}\Big).\label{00}
\eea
Here we have assumed that $ \partial \mathcal{M} $ contains two space-like $(D-1)$-dimensional hyper-surfaces $\Sigma_{t}$ at $ t=\text{constant} $ ($\omega=t$) and one time-like hyper-surface $\mathcal{B}$ at $ r=\text{constant} $ ($\omega=r$). We further choose that the space-time is flat asymptotically at $r=\infty$. (Our arguments also apply for asymptotically anti-de sitter or de Sitter space-times. However, for lexical simplicity, we shall only mention the Minkowski flat space-time.) To obtain \eqref{00}, we have used the definition of canonical momentum in Eq.~(\ref{momenta}). Note also
\begin{equation}
h_{ab(\Sigma)} \pi^{ab(k)}_{(\Sigma)}=h_{ab(\Sigma)} \frac{\partial (N\sqrt{h} L^{(k)}_{{\text{ADM}}})}{\partial (\partial_t h_{ab})}= -\frac{D-2k}{2} \mathcal{L}_{\rm surf (\Sigma)}^{(k)},
\label{33}
\end{equation}
\begin{equation}
h_{ab(\mathcal{B})} \pi^{ab(k)}_{(\mathcal{B})}=h_{ab(\mathcal{B})} \frac{\partial (N\sqrt{h} L^{(k)}_{{\text{ADM}}})}{\partial (\partial_r h_{ab})}= -\frac{D-2k}{2} \mathcal{L}_{\rm surf(\mathcal{B})}^{(k)}.
	\label{34}
\end{equation}
In our relations the indices $(\Sigma)$ and $(\mathcal{B})$ specify the induced metric and extrinsic curvature $K$ and $\bar{R}$ on hyper-surfaces $\Sigma$ and $\mathcal{B}$ respectively.

\subsubsection{Dirichlet BC on $\mathcal{B}$}

Following from Eq.~\eqref{00}, it appears that both Dirichlet BC,  $ \delta h_{ab} \lvert_{\Sigma_{t}, \mathcal{B}}=0 $, and Neumann BC,  $\delta \mathit{\pi}^{ab(k)} \lvert_{\Sigma_{t},\mathcal{B}}=0$, are required in order to obtain the equations of motion {\it via} the variation principle. However, as in the case of Einstein gravity, we can deal with applying only the Dirichlet BC provided with additional appropriate Myers term to the Lovelock action. We shall show that when the Lagrangian is degenerate, the alternative boundary condition is also possible. However, the discussion of the BCs and Myers terms is different, depending on whether the boundary is $\mathcal{B}$ and $\Sigma_{t}$.

We shall first consider the time-like boundary $\mathcal{B}$.  {\it A priori}, we can impose either the Dirichlet or the Neumann BCs.  However, we would like to require that our space-times be asymptotically flat, {\it i.e.}~$h_{ab(\mathcal{B})}=\eta_{ab}$ at the boundary $r=\infty$. This restriction automatically imposes the Dirichlet BC.  The consistency then requires the same BC for all time-like boundaries $\mathcal{B}$. {It should be noticed that when the boundary is taken to be literally at the spatial infinity, the integrals over $\mathcal{B}$ vanish and play no role in the variation principle. Therefore to obtain equation of motion, we can omit the surface integral terms on $\mathcal{B}$ in Eq.(\ref{00}). However if we require that the variational principle be well posed quasilocally (at finite $r$) then we should retain the boundary terms on the lateral boundary $\mathcal{B}$.

Focusing on $\mathcal{B}$, it follows from (\ref{00}) that one can append the Lovelock action and obtain \cite{Pap, Pap1}
\begin{equation}
	\mathcal{A}_{\rm D}^{(k)}=\int_{\mathcal{M}} d^D x\sqrt{-g}L_{\textit{ \rm Lovelock}}^{(k)}+\frac{2}{D-2k}\int_{\mathcal{B}} d^{D-1}yh_{ab(\mathcal{B})}\mathit{\pi}^{ab(k)}_{(\mathcal{B})},
	\label{1}
\end{equation}
where the second term on the right hand side is the required Myers term corresponding to Dirichlet BC. Note that we use subscript D to denote the action that involves the Dirichlet BC and we should not confuse it with dimension $D$. Varying $  \mathcal{A}_{\rm D}^{(k)}$ gives \cite{Pap2}
\begin{equation}
	\delta \mathcal{A}_{\rm D}^{(k)} =\int_{\mathcal{M}} d^D x\sqrt{-g}E^{\mu\nu}\delta g_{\mu\nu}+\int_{\mathcal{B}} d^{D-1}y\mathit{\pi}^{ab(k)}_{(\mathcal{B})}\delta h_{ab(\mathcal{B})}.
\end{equation}
Applying $ \delta h_{ab}|_{\mathcal{B}}=0 $
leads to equation of motion $ E^{\mu\nu}=0 $, where $ E^{\mu\nu} $ is defined in Eq.~(\ref{eom}).

Inserting Eq.~(\ref{moment}) into (\ref{1}), we find
\begin{equation}
	\mathcal{A}_{\rm D}^{(k)}=\int d^{D}x\sqrt{-g}L_{\text{ Lovelock}}^{(k)}-\int_{\mathcal{B}}d^{D-1}y\mathcal{L}_{{\rm surf}(\mathcal{B})}^{(k)}.
\end{equation}
Evaluating this for $k = 1$ gives us the full EH action, consistent with Dirichlet BC, as follows:
\begin{equation}
	\mathcal{A}_{\rm D}^{(1)}=\int_{\mathcal{M}} d^D x\sqrt{-g}L_{\rm EH}-2\int_{\mathcal{B}}d^{D-1}y\sqrt{h}\mathit{K}_{(\mathcal{B})},
	\label{aa}
\end{equation}
where we used $\mathcal{L}_{(\mathcal{B})}^{(1)}=2\sqrt{h}K_{(\mathcal{B})}$. It is clear that the required Myers term on $\mathcal{B}$, compatible with Dirichlet BC, does not depend on the dimensions of space-time and only depends on the induced metric and its time  Lie derivative through $K$.

\subsubsection{Dirichlet BC or Neumann BC on $\Sigma_{t}$}

On the spacelike boundaries $\Sigma_{t}$, the situation is different.  There can exist different choices of the BCs. Here for simplicity we consider asymptotically flat space-time at $r\rightarrow \infty$ to omit the surface integral terms on $\mathcal{B}$.  We can adopt the traditional approach and impose the Dirichlet BC, in which case, the Myers term is the same, leading to
\bea
	\mathcal{A}_{\rm D}^{(k)}&=&\int_{\mathcal{M}} d^D x\sqrt{-g}L_{\rm { Lovelock}}^{(k)}+\frac{2}{D-2k}\int_{\Sigma_{t}} d^{D-1}yh_{ab(\Sigma)}\mathit{\pi}^{ab(k)}_{(\Sigma)}\nn\\
&=&\int_{\mathcal{M}} d^{D}x\sqrt{-g}L_{\rm { Lovelock}}^{(k)}-\int_{\Sigma_t}d^{D-1}y\mathcal{L}_{{\rm surf}(\Sigma)}^{(k)}\,.
	\label{final}
\eea

Alternatively, we can impose the Neumann BC $\delta \pi^{ab(k)}\lvert_{\Sigma_{t}}=0 $.\footnote{In the standard approach, Neumann BC means to fix the first-order normal derivative of a field. For many simple models, this is equivalent to the fixation of the momentum field. In general relativity, fixing the normal derivative of the metric does not lead to the fixation of its momentum. The latter requires to fix some specific linear combination of both the metric and its normal derivative. Therefore, one may want to call this generalization as the ``Neumann-like BC'' or ``Robin BC''. The advantage is to make it manifest that the canonical pair with nonvanishing Poisson bracket would not be fixed simultaneously on the boundary.} This is consistent with the Dirichlet BC on $\mathcal{B}$, since the hyper-surfaces $\Sigma_{t}$ intersect $\mathcal{B}$ orthogonally and hence the BCs on $\Sigma_{t}$ and $\mathcal{B}$ are independent. It follows from  Eq.~(\ref{00}) that we propose the following action
\begin{equation}
	\mathcal{A}_{\rm N}^{(k)}=\int_{\mathcal{M}} d^D x\sqrt{-g}L_{\rm { Lovelock}}^{(k)}-\frac{D-2(k+1)}{D-2k}\int_{\Sigma_t} d^{D-1}y \mathit{\pi}^{ab(k)}_{(\Sigma)}h_{ab(\Sigma)}.
	\label{h01}
\end{equation}
Here we use subscript N to denote the action that involves the Neumann BC. Varying this action, we have
\begin{equation}
	\delta \mathcal{A}_{\rm N}^{(k)}=\int_{\mathcal{M}} d^DxE^{ab}\delta g_{ab} -\int_{\Sigma_t} d^{D-1} yh_{ab(\Sigma)}\delta\mathit{\pi}^{ab(k)}_{(\Sigma)}.
\end{equation}
By imposing the Neumann BC $ \delta\mathit{\pi}^{ab(k)}|_{\Sigma_t}=0 $, we restore the equation of motion $\mathit{E}^{\mu\nu}=0 $.

Let us write the Myers term (\ref{h01}) in a more familiar form using (\ref{moment}), we get
\begin{equation}
	\mathcal{A}^{(k)}_{\rm N}=\int d^{D}x\sqrt{-g}L_{\rm{ Lovelock}}^{(k)}+\frac{(D-2(k+1))}{2}\int_{\Sigma_t}d^{D-1}y \mathcal{L}_{{\rm surf}(\Sigma)}^{(k)}.
	\label{16s}
\end{equation}
For the case $ k=1 $, we have \cite{Krishnan}
\begin{equation}
	\mathcal{A}^{(1)}_{\rm N}=\int d^{D}x\sqrt{-g}L_{\rm{ EH}}+(D-4)\int_{\Sigma_t}d^{D-1}y\sqrt{h}\mathit{K}_{(\Sigma)}.
\end{equation}
Thus, in contrast to the Dirichlet case, the required Myers term compatible with Neumann BC, depends on the dimensions of the space-time. In particular, when $ D=2(k+1) $, the Myers term vanishes. This observation explicitly shows that the general Lovelock action is naturally compatible with Neumann BC for asymptotically flat space-times; as can also be  seen from (\ref{16s}) for $ D=2(k+1) $. Another interesting feature is the relation between Dirichlet Myers term and Neumann one. Comparing equations. (\ref{final}) and (\ref{16s}), we find
\begin{equation}
	\mathcal{A}_{\rm N}^{{\rm{ Myers}}(k)}=-\frac{D-2(k+1)}{2}\mathcal{A}_{\rm D}^{{\rm{ Myers}}(k)}.
\end{equation}

\subsection{Generalizing to $ f(L_{\rm Lovelock}^{(k)})$-gravity}\label{4}

We have so far shown that the Lagrangian of a general Lovelock gravity is classical holographic-degenerate in the ADM formalism.  As we shall see presently, the general $f(\mathcal{L}_{\rm Lovelock}^{(k)})$ Lagrangian is not degenerate. If we want to investigate different BCs in a consistent way, we need to define the momenta conjugate to the field variables in order to distinguish Dirichlet and Neumann BCs where the momentum fields vanish on the boundary. Following the Ostrogradsky approach \cite{Os}, we are able to write the $f(\mathcal{L}_{Lovelock}^{(k)})$ Lagrangian as a degenerate one. To do so, we adopt the scalar-tensor formulation, by introducing a scalar field $\phi$, and write $f(\mathcal{L}_{\rm Lovelock}^{(k)})$ action in the Jordan frame:
\be
\mathcal{A}^{(k)}_{f}=\int_{\mathcal{M}}d^D \sqrt{-g} f(L_{\rm Lovelock}^{(k)})=\int_{\mathcal{M}}d^Dx\sqrt{-g} \big(\phi L_{\rm Lovelock}^{(k)}-V(\phi)\big),
\label{act0}
\ee
in which $\phi= f'(L_{\rm Lovelock}^{(k)})=\partial f(L_{\rm Lovelock}^{(k)})/\partial \mathcal{L}_{\rm Lovelock}^{(k)} $, $V(\phi)=L_{\rm Lovelock}^{(k)}(\phi)\phi- f(L_{\rm Lovelock}^{(k)}(\phi))$ and we assume that $f''(L_{\rm Lovelock}^{(k)}) \neq 0$ \cite{Cap,Sot, Far}. Now using the Eq.~(\ref{hol}) we have
\begin{align}
	\mathit{\mathcal{A}^{(k)}_f}&=\int_{\mathcal{M}}d^Dx \sqrt{-g}\left( \phi L_{\rm { ADM}}^{(k)}-V(\phi)\right)
	\nonumber\\
	&+\frac{2}{D-2k} \int_{\mathcal{M}}d^Dx \phi\, \partial_w \left(-h_{ab}\frac{\partial(\sqrt{-g}L_{\rm{ ADM}}^{(k)})}{\partial(\partial_{w} h_{ab})}\right).
	\label{4747}
\end{align}
The first integral contains only the metric, its first order derivatives and the scalar field $ \phi $. Integrating by parts, the above Lagrangian is degenerate as follows
\begin{align}
	\mathit{\mathcal{A}^{(k)}_f}&=\int_{\mathcal{M}}d^Dx \sqrt{-g}\left( \phi L_{\rm{ ADM}}^{(k)}-V(\phi)\right)
	+\frac{2}{D-2k}\int_{\mathcal{M}}d^Dx \partial_\omega \phi h_{ab} \mathit{M}^{\omega ab(k)}\nonumber\\
	&-\frac{2}{D-2k}\int_{\Sigma_t}d^{D-1}y \phi h_{ab(\Sigma)} \mathit{\pi}^{ab(k)}_{(\Sigma)},
	\label{hb1}
\end{align}
where
\begin{eqnarray}
	\partial_\omega \phi h_{ab} \mathit{M}^{\omega ab(k)}&&\equiv\partial_\omega \phi h_{ab} \frac{\partial(\sqrt{-g}L_{\rm { ADM}}^{(k)})}{\partial(\partial_{\omega} h_{ab})}=-\frac{D-2k}{2}\partial_\omega \phi \mathcal{L}_{\rm surf}^{(k)}\,.
	\label{nhf}
\end{eqnarray}
Note that $ \mathit{\pi}^{ab(k)}_{(\Sigma)}\equiv \partial(\sqrt{-g}L_{\rm{ ADM}}^{(k)})/\partial(\partial_{w} h_{ab})$ is the canonical momentum of $ h_{ab(\Sigma)} $ in Lovelock gravity. Also we have assumed that $ \partial \mathcal{M} $ contains two spacelike $(D-1)$-dimensional surfaces $ \Sigma_t $ and one timelike surface $\mathcal{B}$ on which the integral vanishes at large spatial distances for asymptotically flat space-times. Now we are able to define the canonical momenta of $ \phi $ and $ h_{ab(\Sigma)} $ in $f(L_{Lovelock}^{(k)})$-gravity as
\begin{equation}
	\mathit{\bar{\pi}}^{ab(k)}_{(\Sigma)} \equiv \frac{\delta\mathit{\mathcal{A}^{(k)}_{f}}}{\delta (\partial_t h_{ab})}=\phi \mathit{\pi}^{ab(k)}_{(\Sigma)}+H^{ab(k)}_{(\Sigma)},
	\label{q0}
\end{equation}
where
\bea
H^{ab(k)}_{(\Sigma)}&\equiv&\frac{2}{D-2k} \partial_t\phi g_{de} \frac{\partial \mathit{M}^{t de(k)}}{\partial (\partial_t h_{ab})},\label{H}\\
	\mathit{\bar{\pi}^{(k)}}_{\phi(\Sigma)} &\equiv& \frac{\delta\mathit{\mathcal{A}^{(k)}_{f}}}{\delta (\partial_t \phi)}=\frac{2}{D-2k}\mathit{\pi}^{(k)}_{(\Sigma)}\,,
	\label{q1}
\eea
and  $\mathit{\pi}^{(k)}_{(\Sigma)}=h_{ab(\Sigma)}\mathit{\pi}^{ab(k)}_{(\Sigma)} $.

We now write a classical holographic-Like relation for $f(L_{\rm Lovelock}^{(k)})$-gravity, as Eq.~(\ref{hol}) for general Lovelock theories, to show that, without using Ostrogradsky approach and the scalar-tensor formulation, its Lagrangian is not degenerate. Using Eqs.~(\ref{hb1}), (\ref{q0}), (\ref{nhf}) and $\phi=f'(L_{\rm Lovelock}^{(k)})$, we can obtain
\be
\sqrt{-g} f(L_{\rm Lovelock}^{(k)})=\sqrt{-g} L^{(k)}_{\rm bulk}+\frac{2}{D-2k}\partial_\omega \big(-h_{ab} \frac{\partial(\sqrt{-g} L^{(k)}_{\rm bulk})}{\partial_\omega h_{ab}}+ h_{ab}H^{ab(k)}\big),
\label{hn0}
\ee
in which we have
\be
L_{\rm bulk}^{(k)}=f'(L_{\rm Lovelock}^{(k)}) (L_{ADM}^{(k)}- L_{\rm Lovelock}^{(k)})-f(L_{\rm Lovelock}^{(k)})-\partial_\omega \phi \mathcal{L}_{\rm surf}^{(k)}\,.
\ee
Considering Eq.~(\ref{hn0}), we see that the surface part of the Lagrangian is not determined completely by its bulk part. This is in contrast to the Lovelock Lagrangian. Furthermore, the bulk Lagrangian in $f(L_{\rm Lovelock}^{(k)})$-gravity is not necessarily a first-order Lagrangian and contains an arbitrary function of the second order derivatives of metric. Hence, as we mentioned, the $f(L_{\rm Lovelock}^{(k)})$ Lagrangian is not a degenerate one.

Now varying the degenerated action in Eq.~(\ref{hb1}) with respect to $ \phi $ and $ g_{ab}$, after a little algebra, we obtain
\begin{equation}
	\delta \mathit{\mathcal{A}^{(k)}_{f}}= \delta_{\phi}\mathit{\mathcal{A}^{(k)}_{f}}+\delta_{h}\mathit{\mathcal{A}^{(k)}_{f}},
	\label{9a}
\end{equation}
where
\begin{equation}
	\delta_{\phi}\mathit{\mathcal{A}^{(k)}_{f}}=\int_{\mathcal{M}}d^Dx L_{\phi}^{(k)}\delta \phi,
	\label{8a}
\end{equation}
$$L_{\phi}\equiv \left\{\sqrt{-g}\left( L_{\text{ ADM}}^{(k)}-\partial_\phi V(\phi)\right) +\frac{2}{D-2k} \partial_\omega \left(-h_{ab}\frac{\partial(\sqrt{-g}L_{\text{ ADM}}^{(k)})}{\partial(\partial_{\omega} h_{ab})}\right)\right\},$$
and
\begin{align}
	\delta_{h}\mathit{\mathcal{A}^{(k)}_{f}}&=\int_{\mathcal{M}}d^Dx \mathit{L}^{ab(k)} \delta h_{ab} + \int_{\Sigma{t}}d^{D-1}y \mathit{\bar{\pi}}^{ab(k)}_{(\Sigma)} \delta h_{ab(\Sigma)}-\int_{\Sigma_t}d^{D-1}y \phi \delta \mathit{\bar{\pi}^{(k)}}_{\phi(\Sigma)}
	\label{5353} ,
\end{align}
in which
\begin{align}
	\mathit{L}^{ab(k)}&=\phi \frac{ \partial(\sqrt{-g}L_{\text{ ADM}}^{(k)})}{\partial h_{ab}} - \partial _\omega \left(\phi \mathit{M}^{\omega ab}\right)-\frac{1}{2}\sqrt{-g} h^{ab} V(\phi)+\frac{2}{D-2k}\bigg( \partial_\omega \phi \mathit{M}^{\omega ab(k)} \nonumber \\
	&+\partial_\omega \phi h_{kl}\mathit{B}^{\omega abkl(k)}- \partial_c (\partial_\omega \phi h_{ab} \tilde{B}^{\omega abcde(k)})\bigg),
\end{align}
and $\mathit{B}^{\omega abkl(k)} \equiv \partial\mathit{M}^{\omega ab(k)}/\partial g_{kl} $, $\tilde{B}^{\omega abcde(k)}\equiv \partial\mathit{M}^{\omega ab(k)}/\partial (\partial_c g_{de})$. Substituting (\ref{8a}) and (\ref{5353}) into (\ref{9a}) gives
\begin{align}
	\delta \mathit{\mathcal{A}^{(k)}_{f}}&=\int_{\mathcal{M}}d^Dx L_{\phi}^{(k)}\delta \phi+ \int_{\mathcal{M}}d^Dx \mathit{L}^{ab(k)} \delta h_{ab} + \int_{\Sigma_{t}}d^{D-1}y \mathit{\bar{\pi}}^{ab(k)}_{(\Sigma)} \delta h_{ab(\Sigma)}
	\nonumber\\&
	-\int_{\Sigma_t}d^{D-1}y \phi \delta \mathit{\bar{\pi}^{(k)}}_{\phi(\Sigma)}\,.
	\label{vari}
\end{align}
As in Lovelock gravities discussed earlier, we shall impose only the Dirichlet BC on $\mathcal{B}$. However, based on \eqref{vari}, we find that there exist four different types of BCs on the hyper-surfaces $\Sigma_t$ with appropriate Myers terms that all lead to the well-defined variation principle for $f (L_{\rm Lovelock}^{(k)})$-gravity.  We shall therefore discuss these BCs on $\Sigma_t$ next.

\subsubsection{Dirichlet BC}

In order to impose solely the Dirichlet BC: $ \delta h_{ab}|_{\Sigma_t}=\delta \phi|_{\Sigma_t}=0 $ to the Eq.~(\ref{vari}), we need to modify the action (\ref{act0})  by adding the following Myers term
\begin{align}
	\mathcal{A}^{\text{ D}(k)}_{f}&=	\mathcal{A}^{(k)}_{f}+	\mathcal{A}^{\text{ Myers(D)}(k)}_{f}=\mathcal{A}^{(k)}_{f}+\int_{\Sigma_t}d^{D-1}y \phi \mathit{\bar{\pi}^{(k)}}_{\phi(\Sigma)}
	\label{mn0}
\end{align}
Varying the above action shows that it is compatible with Dirichlet BC as follows
\begin{align}
	\delta \mathcal{A}^{\text{ D}(k)}_{f}&= \int_{\mathcal{M}}d^Dx L_{\phi}^{(k)}\delta \phi+\int_{\mathcal{M}}d^Dx \mathit{L}^{ab(k)} \delta h_{ab} + \int_{\Sigma_t}d^{D-1}y \mathit{\bar{\pi}}^{ab(k)}_{(\Sigma)}  \delta h_{ab(\Sigma)} \nonumber \\
	&+\int_{\Sigma_t}d^{D-1}y \mathit{\bar{\pi}^{(k)}}_{\phi(\Sigma)} \delta \phi\,,
	\label{com}
\end{align}
which gives the equation of motion subjected to Dirichlet BC. Also we can determine the Myers term $\mathcal{A}^{\text{ Myers(D)}(k)}_{f}$ in terms $\mathcal{L}_{\rm surf}^{(k)}$. At first using (\ref{q1}) we can rewrite the Myers term in Eq.~(\ref{mn0}) as
\be
\mathcal{A}^{\text{ Myers(D)}(k)}_{f}=\frac{2}{D-2k}\int_{\Sigma_t}d^{D-1}y \phi h_{ab(\Sigma)} \mathit{\pi}^{ab(k)}_{(\Sigma)}.
\ee
Now substituting $ \phi= f'(L_{\rm Lovelock}^{(k)}) $ and using (\ref{moment}), one can obtain
\be
\mathcal{A}^{\text{ Myers(D)}(k)}_{f}=-\int_{\Sigma_t}d^{D-1}y  f'(L_{\rm Lovelock}^{(k)}) \mathcal{L}_{{\rm surf}(\Sigma)}^{(k)}.
\label{cod}
\ee
Substituting $k=1$ in the above equation one can obtain
\be
\mathcal{A}^{\text{ Myers(D)}}_{f(R)}=-2\int_{\Sigma_t}d^{D-1}y \sqrt{h} f'(R) K_{(\Sigma)}\,,
\ee
which reproduces the result in \cite{HKH1}.

\subsubsection{Neumann BC}

We now consider Neumann BC: $\delta \mathit{\bar{\pi}}^{ab(k)}|_{\Sigma_t}=\delta \mathit{\bar{\pi}^{(k)}}_\phi|_{\Sigma_t}=0 $, and obtain the appropriate Myers term. We propose
\be
\mathcal{A}^{\text{ N}(k)}_{f}=	\mathcal{A}^{(k)}_{f}+	\mathcal{A}^{\text{ Myers(N)}(k)}_{f}=\mathcal{A}^{(k)}_{f}-\int_{\Sigma_t}d^{D-1}y h_{ab(\Sigma)} \mathit{\bar{\pi}}^{ab(k)}_{(\Sigma)},
\label{N}
\ee
Variation of (\ref{N}) yields
\begin{align}
	\delta \mathcal{A}^{\text{ N}(k)}_{f}&= \int_{\mathcal{M}}d^Dx L_{\phi}^{(k)}\delta \phi+\int_{\mathcal{M}}d^Dx \mathit{L}^{ab(k)} \delta h_{ab} - \int_{\Sigma_t}d^{D-1}yh_{ab(\Sigma)}  \delta \mathit{\bar{\pi}}^{ab(k)}_{(\Sigma)}  \nonumber \\
	&+\int_{\Sigma_t}d^{D-1}y \phi\delta\mathit{\bar{\pi}^{(k)}}_{\phi(\Sigma)}\,,
	\label{com1}
\end{align}
which gives the equations of motion using Neumann BC. Using Eqs.~(\ref{q0}) we can rewrite the Myers term so that
\be
\mathcal{A}^{\text{ Myers(N)}(k)}_{f}=-\int_{\Sigma_t}d^{D-1}y \phi h_{ab(\Sigma)} \mathit{\pi}^{ab(k)}_{(\Sigma)} -\int_{\Sigma_t}d^{D-1}y h_{ab(\Sigma)} H^{ab(k)}_{(\Sigma)},
\ee
and using (\ref{moment}) and inserting $ \phi=f'(L_{\rm Lovelock}^{(k)})$, we have the above formula as
\be
\mathcal{A}^{\text{ Myers(N)}(k)}_{f}=\frac{D-2k}{2}\int_{\Sigma_t}d^{D-1}y f'(L_{\rm Lovelock}^{(k)}) \mathcal{L}_{{\rm surf}(\Sigma)}^{(k)}-\int_{\Sigma_t}d^{D-1}y H^{(k)}_{(\Sigma)},
\label{con}
\ee
It should be noted that unlike the case of Dirichlet BC, Neumann Myers term depends on the dimensions of space-time. Also substituting $k=1$ in Eq.~(\ref{H}) gives
\be
H^{(k)}_{(\Sigma)}=N\sqrt{|h|} \frac{2(D-1)}{D-2} \partial_t f'(R),
\ee
so we have
\be
\mathcal{A}^{\text{ Myers(N)}}_{f(R)}=(D-2)\int_{\Sigma_t}d^{D-1}y\sqrt{|h|} f'(R) K_{(\Sigma)}-(D-1)\int_{\Sigma_t}d^{D-1}y N\sqrt{|h|} \partial_t f'(R),
\ee
which is compatible with the result in \cite{HKH1}. It is worth to compare (\ref{cod}) with the Myers term  (\ref{con}) for Dirichlet BC. It is easily seen that
\begin{equation}
	\mathcal{A}^{\text{ Myers(N)}(k)}_{f}=-\frac{D-2k}{2}\mathcal{A}^{\text{ Myers(D)}(k)}_{f}-\int_{\Sigma_t}d^{D-1}y H^{(k)}_{(\Sigma)}\,.
\end{equation}
\subsubsection{Mixed BCs}

There can exist two types of consistent mixed BCs for $f(L_{\rm Lovelock}^{(k)})$-gravity: $ \delta\mathit{\bar{\pi}}^{ab(k)}|_{\Sigma_t}=\delta \phi|_{\Sigma_t}=0$, or $ \delta \mathit{\bar{\pi}^{(k)}}_\phi|_{\Sigma_t}=\delta h_{ab}|_{\Sigma_t}=0$. We begin with the first one. Using the variation of $f(L_{\rm Lovelock}^{(k)})$-gravity action (\ref{vari}), the first type mixed BC is consistent if we add the appropriate Myers term so that the full action is
\be
	\mathcal{A}^{\text{ M1}(k)}_{f}=	\mathcal{A}^{(k)}_{f}+	\mathcal{A}^{\text{ Myers(M1)}(k)}_{f}=\mathcal{A}^{(k)}_{f}-\int_{\Sigma_t}d^{D-1}y h_{ab(\Sigma)} \mathit{\bar{\pi}}^{ab(k)}_{(\Sigma)}
	+\int_{\Sigma_t}d^{D-1}y \phi \mathit{\bar{\pi}^{(k)}}_{\phi(\Sigma)},
	\label{m1}
\ee
Varying this action gives
\begin{align}
	\delta \mathcal{A}^{\text{ M1}(k)}_{f}=& \int_{\mathcal{M}}d^Dx L_{\phi}^{(k)}\delta \phi+\int_{\mathcal{M}}d^Dx \mathit{L}^{ab(k)} \delta g_{ab} - \int_{\Sigma_t}d^{D-1}yh_{ab(\Sigma)}  \delta \mathit{\bar{\pi}}^{ab(k)}_{(\Sigma)}  \nonumber \\
	&+\int_{\Sigma_t}d^{D-1}y \mathit{\bar{\pi}^{(k)}}_{\phi(\Sigma)}\delta \phi\,.
	\label{comi1}
\end{align}
Therefore, the first mixed BC indeed yields the equations of motion. Using Eqs.~(\ref{q0}) and (\ref{q1}), the Myers term in (\ref{m1}) can be written as
\be
\mathcal{A}^{\text{ Myers(M1)}(k)}_{f}=-\frac{D-2(k+1)}{D-2k}\int_{\Sigma_t}d^{D-1}y \phi h_{ab(\Sigma)} \mathit{\pi}^{ab(k)}_{(\Sigma)}-\int_{\Sigma_t}d^{D-1}y h_{ab(\Sigma)} H^{ab(k)}_{(\Sigma)}.
\label{z0}
\ee
Now using Eq.~(\ref{moment}) and $ \phi=f'(L_{\rm Lovelock}^{(k)}) $, we can write Myers term of Eq.~(\ref{z0}) in terms of $\mathcal{L}^{(k)}_{\rm surf}$ as
\be
\mathcal{A}^{\text{ Myers(M1)}(k)}_{f}=\frac{D-2(k+1)}{2}\int_{\Sigma_t}d^{D-1}y f'(\mathcal{L}_{\rm Lovelock}^{(k)}) \mathcal{L}_{{\rm surf}(\Sigma)}^{(k)}-\int_{\Sigma_t}d^{D-1}y H^{(k)}_{(\Sigma)}.
\label{c1}
\ee
Choosing $k=1$ gives us
\be
\mathcal{A}^{\text{Myers(M1)}}_{f(R)}=(D-4)\int_{\Sigma_t}d^{D-1}y\sqrt{h} f'(R) K_{(\Sigma)}-(D-1)\int_{\Sigma_t}d^{D-1}y N\sqrt{h} \partial_t f'(R),
\ee
which is also compatible with the result in \cite{HKH1}. It is also worth comparing the Dirichlet and the above mixed Myers boundary terms in $ f(L_{\rm Lovelock}^{(k)}) $-gravity. We see
\begin{equation}
	\mathcal{A}^{\text{ Myers(M1)}(k)}_{f}=-\frac{D-2(k+1)}{2}\mathcal{A}^{\text{ Myers(D)}(k)}_{f}-\int_{\Sigma_t}d^{D-1}y H^{(k)}_{(\Sigma)}.
\end{equation}

Finally we turn to the second type of mixed BC. Considering Eq.~(\ref{vari}) it is quite clear that by applying this BC, we can get the equations of motion without adding any Myers terms to the action. This means that $ f(L_{\rm Lovelock}^{(k)}) $-gravity with this mixed BC is self-consistent with no need to any Myers term in all $D$ dimension.

\subsection{Physical quantities and Myers terms}\label{5}

Although the equations of motion are the same for imposing different BCs, the total actions are different because of the different Myers terms.  It is thus important to the address the physical implications since the actions play an important role in quantum gravity. Here we consider black hole thermodynamics in the semiclassical approximation in the path integral approach \cite{Hint,HKH2,Brown}. The partition function in the semiclassical limit and for an arbitrary gravitational model takes the form
\begin{equation}
	\mathcal{Z}=\int [dg] e^{-\tilde{\mathcal{A}^*_E}} \simeq e^{-\tilde{\mathcal{A}^*_E}},
	\label{eq17}
\end{equation}
where $\mathcal{A}^*_E=\mathcal{A}_E-\mathcal{A}_{E0}$ in which $\mathcal{A}_E $ is the Euclidean action and $\mathcal{A}_{E0} $ is the corresponding  background action. The symbol tilde means dividing by $16\pi G$, {\it i.e.}~$\tilde{\mathcal{A}^*_E}= \lim_{r\rightarrow \infty} \frac{\mathcal{A}^*_E}{16\pi G}$. The free energy, entropy, and energy are then
\be
	F=\frac{-1}{\beta} ln \mathcal{Z}=\frac{1}{\beta}\tilde{\mathcal{A}^*_E},\qquad
	E=F+\beta\frac{\partial F}{\partial \beta},\qquad
	S=\beta^2 \frac{\partial F}{\partial \beta }.
	\label{1234}
\ee
It could be troubling if different actions based on different BCs lead to different black hole entropies, since we certainly do not expect they would all satisfy the same first law of black hole thermodynamics. Since we need to obtain the numerical value of the Euclidean action for a black hole in the region $r_{H}<r<r_{\infty}$, we should first make the action well defined in the region where the manifold $\mathcal{B}$ is considered for large but finite $r$. In this case the surface terms on the manifold $\mathcal{B}$ no more vanish naturally. (They vanish only at $r \rightarrow \infty$ for asymptotically flat space-times \cite{HKH2}.) Hence, in order to make the Lovelock action well defined one should consider the following two actions:

\bea
	\mathcal{A}^{(k)}_{\rm D}&=& \int_\mathcal{M} d^Dx\sqrt{-g} L_{\text{ Lovelock}}^{(k)}+ \frac{2}{D-2k} \int_{\mathcal{B}} d^{D-1}z h_{ab(\mathcal{B})} \mathit{\pi}^{ab(k)}_{(\mathcal{B})}\nn\\
&&+ \frac{2}{D-2k} \int_{\Sigma_{t}} d^{D-1}y h_{ab(\Sigma)} \mathit{\pi}^{ab(k)}_{(\Sigma)},\nn\\
\mathcal{A}^{(k)}_{\rm N}&=&\int_\mathcal{M} d^Dx\sqrt{-g}L_{\text{ Lovelock}}^{(k)}+ \frac{2}{D-2k} \int_{\mathcal{B}} d^{D-1}z h_{ab(\mathcal{B})} \mathit{\pi}^{ab(k)}_{(\mathcal{B})}\nn\\
&&+\frac{D-2(k+1)}{D-2k} \int_{\Sigma_{t}} d^{D-1}y \mathit{\pi}^{ab(k)}_{(\Sigma)} h_{ab(\Sigma)}.
\eea
The situation is analogous for the well defined $f(L_{\rm Lovelock}^{(k)})$-gravity actions that allow four different boundary conditions on $\Sigma_t$. The full actions associated with these four types BCs are:
\bea
	\mathcal{A}^{\text{ D}(k)}_{f} &=&\mathcal{A}^{(k)}_{f} +\int_{\mathcal{B}}d^{D-1}y \phi \mathit{\bar{\pi}^{(k)}}_{\phi(\mathcal{B})}+\int_{\Sigma_t}d^{D-1}y \phi \mathit{\bar{\pi}^{(k)}}_{\phi(\Sigma)},\nn\\
\mathcal{A}^{\text{ N}(k)}_{f} &=&\mathcal{A}^{(k)}_{f}+\int_{\mathcal{B}}d^{D-1}y \phi \mathit{\bar{\pi}^{(k)}}_{\phi(\mathcal{B})}-\int_{\Sigma_t}d^{D-1}y h_{ab(\Sigma)}\mathit{\bar{\pi}}^{ab(k)}_{(\Sigma)},\nn\\
\mathcal{A}^{\text{ M1}(k)}_{f}&=&\mathcal{A}^{(k)}_{f} +\int_{\mathcal{B}}d^{D-1}y \phi \mathit{\bar{\pi}^{(k)}}_{\phi(\mathcal{B})} -\int_{\Sigma_t}d^{D-1}y h_{ab(\Sigma)} \mathit{\bar{\pi}}^{ab(k)}_{(\Sigma)} +\int_{\Sigma_t}d^{D-1}y \phi \mathit{\bar{\pi}^{(k)}}_{\phi(\Sigma)},\nn\\
\mathcal{A}^{\text{ M2}(k)}_{f}&=&\mathcal{A}^{(k)}_{f}+\int_{\mathcal{B}}d^{D-1}y \phi \mathit{\bar{\pi}^{(k)}}_{\phi(\mathcal{B})}.
\eea
As was explained previously, the Myers boundary terms are the same on $\mathcal{B}$. The difference lies on the boundary $\Sigma_t$. For Euclidean actions, we do not have $\Sigma_t$ and hence the issue does not arise such as black holes.  In Lorentzian signature, the $\Sigma_t$ boundary terms will not contribute either for any stationary geometries. They can however have nontrivial effects on cosmological backgrounds and it requires further investigation.

\section{A new foliation-independent classical holographic relation}

In the previous section, we showed that in the ADM formalism, the general Lovelock gravity is degenerate and there is a classical holographic relation between the surface and bulk terms. This can be viewed as a generalization of the classical holographic relation in Einstein gravity.  However, there is one big difference. In the ADM formalism, the coordinate $w$ is specially treated as the foliation coordinate while the classical holographic relation for the Einstein gravity can be foliation independent in that there is no coordinate that is special.  It is then natural to ask the question whether such a foliation-independent formulation as Eq.~\eqref{hhh} also exist in Lovelock gravities. In this section, we show that there indeed exists a foliation-independent classical holographic relation, but it is inequivalent from the one in the ADM formalism for $k\ge 2$.

\subsection{$k=1$: Einstein gravity}

We begin with the definition of the Riemann tensor (for torsion-free connections)
\bea
R^{\nu_1\nu_2}_{\mu_1\mu_2} =
g^{\nu_2\alpha} \Big(\partial_{\mu_1} \Gamma^{\nu_1}{}_{\mu_2\alpha} -\partial_{\mu_2} \Gamma^{\nu_1}{}_{\mu_1\alpha}\Big)
+g^{\nu_2\alpha}\Big(\Gamma^{\nu_1}{}_{\mu_1\beta} \Gamma^\beta{}_{\mu_2\alpha}-\Gamma^{\nu_1}{}_{\mu_2\beta} \Gamma^\beta{}_{\mu_1\alpha}\Big).
\eea
The indices $(\mu_1,\mu_2)$ are manifestly antisymmetric. What is nontrivial is that
$(\nu_1,\nu_2)$ in the Riemann tensor is also antisymmetric, even though they are not in each of the two brackets of the right-hand side of the equation. It is thus instructive to define
\be
(\Gamma^2)^{\nu_1\nu_2}_{\mu_1 \mu_2} = \Gamma^{[\nu_1}{}_{\alpha [\mu_1]} \Gamma^{|\alpha|}{}_{\mu_2]\beta} g^{\nu_2]\beta}\,,\qquad
(D\Gamma^2)^{\nu_1\nu_2}{}_{\mu_1\mu_2} = \partial_{[\mu_1} \Gamma^{[\nu_1}{}_{\mu_2]\alpha} g^{\nu_2]\alpha}\,.\label{Gammasq}
\ee
In other words, we force antisymmetrization of $(\nu_1,\nu_2)$ in each term. We can then also express the Riemann tensor as
\be
R^{\nu_1\nu_2}_{\mu_1\mu_2}=2(\Gamma^2)^{\nu_1\nu_2}_{\mu_1 \mu_2} + 2(D\Gamma^2)^{\nu_1\nu_2}_{\mu_1\mu_2}\,.\label{newexpforriemann}
\ee
The Ricci scalar is thus given by
\be
R=\delta^{\mu_1\mu_2}_{\nu_1\nu_2} R^{\nu_1\nu_2}_{\mu_1\mu_2} =L^{(1)}_0 + L^{(1)}_1\,,
\ee
where
\be
L^{(1)}_0 = 2 \delta^{\mu_1\mu_2}_{\nu_1\nu_2} (\Gamma^2)^{\nu_1\nu_2}_{\mu_1 \mu_2}\,,\qquad
L^{(1)}_1 = 2 \delta^{\mu_1\mu_2}_{\nu_1\nu_2} (D\Gamma^2)^{\nu_1\nu_2}_{\mu_1\mu_2}\,.
\ee
It can be established that the combination $\sqrt{-g} (L_1^{(1)} + 2 L_0^{(0)})$ is a total derivative in a foliation-independent way:
\be
{\cal L}^{(1)}_{\rm surf}(\Gamma, \partial\Gamma)=\sqrt{-g} (L_1^{(1)} +2 L_0^{(1)})=\partial_\mu J^{\mu (1)}\,,\qquad
J^{\mu(1)} = \sqrt{-g} \Big(g^{\rho\sigma} \Gamma^{\mu}{}_{\rho\sigma} - g^{\rho\mu} \Gamma^\sigma{}_{\rho\sigma}\Big).\label{GBtwoparts}
\ee
Compare the structure $J^\mu$ and $L^{(1)}_0$, we have the identity
\be
J^{\mu (1)} =\delta^\sigma_\rho \fft{\partial(\sqrt{-g}\, L_0^{(1)})}{\partial \Gamma^\sigma{}
_{\mu\rho}}\,.
\ee
This leads to the classical holographic relation for Einstein gravity
\be
\sqrt{-g} R =\sqrt{-g} L_{\rm Lovelock}^{(1)}(\Gamma, \partial\Gamma)=
\sqrt{-g} L_{\rm bulk}^{(1)}(\Gamma) -\partial_\mu
\Big(\delta^\sigma_\rho \fft{\partial(\sqrt{-g}\, L_{\rm bulk}^{(1)})}{\partial \Gamma^\sigma{}
_{\mu\rho}}\Big),
\ee
where $L_{\rm bulk}^{(1)}=-L_0^{(1)}$.  Making use of the identity
\be
\fft{\partial\Gamma^{\rho}{}_{\mu\nu}}{\partial (\partial_\alpha g_{\beta\gamma})}=
\fft12\Big(-g^{\rho\alpha}\delta^\beta_\mu\delta^\gamma_\nu + g^{\rho\beta} \delta_{\mu}^\alpha\delta_{\nu}^\gamma + g^{\rho\gamma} \delta_{\mu}^\beta\delta_{\nu}^\alpha\Big),\label{Gammaid1}
\ee
the classical holographic relation can also be written as \cite{pad1,pad2}
\be
\sqrt{-g} R = \sqrt{-g} L_{\rm bulk}^{(1)}(\Gamma) -\fft{2}{D-2}\partial_\mu
\Big(g_{\nu\rho} \fft{\partial(\sqrt{-g}\, L_{\rm bulk}^{(1)})}{\partial (\partial_\mu g_{\nu\rho})}\Big)\,.
\ee
It is important to note that for Einstein gravity, the $L_{\rm bulk}^{(1)}(\Gamma)$ obtained in this foliation-independent approach is identically the same as the bulk term in the ADM formalism discussed in the previous section.

\subsection{$k=2$: Gauss-Bonnet gravity}

With our new expression for the Riemann tensor \eqref{newexpforriemann}, it can be easily seen that the Lagrangian of GB gravity is
\be
\sqrt{-g} L_{\rm GB}= \sqrt{-g} \big(L^{(2)}_{0} + 2L^{(2)}_{1} + L^{(2)}_{2}\big).
\ee
where
\bea
L^{(2)}_{0} &=& 4!\, \delta^{\mu_1\mu_2\mu_3\mu_4}_{\nu_1\nu_2\nu_3\nu_4}\, (\Gamma^2)^{\nu_1\nu_2}_{\mu_1\mu_2}\, (\Gamma^2)^{\nu_3\nu_4}_{\mu_3\mu_4}\,,\nn\\
L^{(2)}_{1} &=& 4!\, \delta^{\mu_1\mu_2\mu_3\mu_4}_{\nu_1\nu_2\nu_3\nu_4}\, (\Gamma^2)^{\nu_1\nu_2}_{\mu_1\mu_2}\, (D\Gamma^2)^{\nu_3\nu_4}_{\mu_3\mu_4}\,,\nn\\
L^{(2)}_{2} &=& 4!\, \delta^{\mu_1\mu_2\mu_3\mu_4}_{\nu_1\nu_2\nu_3\nu_4}\, (D\Gamma^2)^{\nu_1\nu_2}_{\mu_1\mu_2}\, (D\Gamma^2)^{\nu_3\nu_4}_{\mu_3\mu_4}\,.
\eea
By some nontrivial exercise that we shall present momentarily, we find that
\be
\sqrt{-g} L_{\rm GB}= \sqrt{-g} L^{(2)}_{\rm bulk} +
\partial_\mu J^{\mu (2)}\,,\label{GBbulksurf}
\ee
where
\be
L^{(2)}_{\rm bulk}(\Gamma,\partial\Gamma)=-L^{(2)}_{0} - L^{(2)}_1\,,\qquad J^{\mu (2)}=4!\,\sqrt{-g}\,\delta^{\mu\mu_2\mu_3\mu_4}_{\nu_1\nu_2\nu_3\nu_4}\, \Gamma^{\nu_1}\!{}_{\mu_2}\!{}^{\nu_2} \,\Big((\Gamma^2)^{\nu_3\nu_4}_{\mu_3\mu_4} + (D\Gamma^2)^{\nu_3\nu_4}_{\mu_3\mu_4}\Big).
\ee
It can be shown that GB gravity is classical holographic, namely
\be
J^{\mu(2)} = - \delta^\nu_\rho \fft{\partial (\sqrt{-g} L^{(2)}_{\rm bulk})}{\partial \Gamma^\nu{}_{\mu\rho}} - \Gamma^\nu{}_{\rho\sigma} \fft{\partial (\sqrt{-g} L^{(2)}_{\rm bulk})}{\partial (\partial_\mu\Gamma^\nu{}_{\rho\sigma})}.\label{GBholorelation}
\ee
In other words, the surface term is completely specified by the bulk action.

   We now give the proof. We begin by expressing the GB term as
\be
L_{\rm GB}=A_{\nu_1 \nu_2}^{\mu_1 \mu_2} R^{\nu_1 \nu_2}_{\mu_1 \mu_2},\qquad
A_{\nu_1 \nu_2}^{\mu_1 \mu_2} \equiv \frac{4!}{2^2}   \delta_{\nu_1 \nu_2 \nu_3 \nu_4}^{\mu_1 \mu_2 \mu_3 \mu_4} R^{\nu_3 \nu_4}_{\mu_3 \mu_4}.
\label{ah0}
\ee
The quantity $A_{\nu_1 \nu_2}^{\mu_1 \mu_2}$ has the symmetries of the Riemann tensor and it is divergence free, namely $\nabla_{\mu_1} A_{\nu_1 \nu_2}^{\mu_1 \mu_2}=0$ \cite{pad1,pad2}. Expressing $R^{\nu_1 \nu_2}_{\mu_1 \mu_2}$ in terms of $\Gamma^{\nu_1}{}_{\nu_2 \mu_1}$, we can write Eq.~(\ref{ah0}) as
\begin{align}
	\sqrt{-g}L_{\rm GB}=&2\sqrt{-g} A_{\nu_1}{}^{\nu_2 \mu_1 \mu_2} (\partial_{\mu_1} \Gamma^{\nu_1}{}_{\mu_2 \nu_2}+\Gamma^{\nu_1}{}_{\mu_1 \mu}\Gamma^\mu{}_{\mu_2 \nu_2})=\nonumber \\
	&2\sqrt{-g}\big( A_{\nu_1}{}^{\nu_2 \mu_1 \mu_2} \Gamma^{\nu_1}{}_{\mu_1 \mu}\Gamma^\mu{}_{\nu_2 \mu_2}- A_{\nu_1}{}^{\nu_2 \mu_1 \mu_2} \Gamma^{\nu_1}{}_{\mu_2\nu_2}\Gamma^\mu{}_{\mu_1 \mu}
	- \Gamma^{\nu_1}{}_{\mu_2 \nu_2} \partial_{\mu_1} A_{\nu_1}{}^{\nu_2 \mu_1 \mu_2} \big)
	\nonumber \\
	&+2\partial_{\mu_1} [ \sqrt{-g} A_{\nu_1}{}^{\nu_2 \mu_1 \mu_2} \Gamma^{\nu_1}{}_{\mu_2 \nu_2} ] .
	\label{ah1}
\end{align}
It follows from the divergence free condition of $A$, we have
\begin{align}
	\partial_{\mu_1} A_{\nu_1}{}^{\nu_2 \mu_1 \mu_2}=- \Gamma^{\nu_2}{}_{\mu \mu_1} A_{\nu_1}{}^{\mu \mu_1 \mu_2}+\Gamma^\mu{}_{\nu_1 \mu_1} A_{\mu}{}^{\nu_2 \mu_1 \mu_2}- \Gamma^{\mu_1}{}_{\mu \mu_1} A_{\nu_1}{}^{\nu_2 \mu \mu_2}.
	\label{ah2}
\end{align}
Substituting this into Eq.~(\ref{ah1}) and writing $A_{\nu_1}{}^{\nu_2 \mu_1 \mu_2}$ in terms of $\Gamma^{\nu_1}{}_{\nu_2 \mu_1}$ and $\delta_{\nu_1 \nu_2 \nu_3 \nu_4}^{\mu_1 \mu_2 \mu_3 \mu_4}$, we obtain \eqref{GBbulksurf}. To obtain the classical holographic relation, we note
\bea
&&\delta ^{\nu_1}_{\nu_2} \frac{\partial(\sqrt{-g} L^{(2)}_{\rm bulk})}{\partial \Gamma^{\nu_1}{}_{\mu_1 \nu_2}}=-4!\sqrt{-g} g^{\nu_2 \nu} g^{\nu_4 \nu'} \Big( \delta_{\nu_1 \nu \nu_3 \nu'} ^{\mu_1 \mu_2 \mu_3 \mu_4}( \Gamma^{\nu_1}{}_{\nu_2 \mu_2} \Gamma^{\nu_3}{}_{\mu_3 \rho}\Gamma^{\rho}{}_{\mu_4 \nu_4}+\Gamma^{\nu_1}{}_{\nu_2 \mu_2} \partial_{\mu_3} \Gamma^{\nu_3}{}_{\mu_4 \nu_4}) \nn\\
&&\qquad\qquad\qquad\qquad\quad+ \delta_{\nu_1 \nu \nu_3 \nu'} ^{ \mu_3 \mu_2 \mu_4 \mu_1} \Gamma^{\nu_1}{}_{\mu_2 \rho} \Gamma^{\rho}{}_{\nu_2 \mu_3}\Gamma^{\nu_3}{}_{\mu_4 \nu_4}  \Big),\nn\\
&&\Gamma^{\nu_1}{} _{\mu_4 \nu_2} \frac{\partial(\sqrt{-g} L^{(2)}_{\rm bulk})}{\partial (\partial_{\mu_1} \Gamma^{\nu_1}{} _{\mu_4 \nu_2})}=4! \sqrt{-g}g^{\nu_2 \nu} g^{\nu_4 \nu'} \delta_{\nu_1 \nu \nu_3 \nu'} ^{\mu_3 \mu_2 \mu_4 \mu_1} \Gamma^{\nu_1}{}_{\mu_2 \rho} \Gamma^{\rho}{}_{\nu_2 \mu_3}\Gamma^{\nu_3}{}_{\mu_4 \nu_4}.
	\label{hn1}
\eea
With these, the classical holographic relation \eqref{GBholorelation} follows straightforwardly.

It is worth remarking that it is nontrivial to demonstrate that $\sqrt{-g} L_{\rm bulk}$ gives the foliation-independent equation of motion \eqref{eom} of $k=2$. We shall illustrate here how a quadratic Riemann tensor theory that might involve four derivatives in the equations of motion actually reduces to a two-derivative theory.  The fact that $L_{\rm bulk}$ does not involve $L^{(2)}_2$ implies that the equation of motion can be at most of three derivatives, originated from $L^{(2)}_1$. However, with a straightforward exercise, the $\sqrt{-g} L^{(2)}_1$ will not contribute three derivatives to the equation of motion, even though it cannot express as a total derivative in a foliation-independent way such as $\partial_\mu J^\mu$. The situation improves if we break the covariance and treat one coordinate special, such as $w$ coordinate in the ADM formalism. The $L^{(2)}_1$ can be further decomposed into a bulk part with no $\partial_w^2$ and a surface term expressed as $\partial_w J$. It is perhaps demonstrative to consider an example of classical mechanics with the analogous Lagrangian
\bea
L_2 &=& q_1 q_2 (\dot q_1 \dot q_2)^m (\ddot q_1 \dot q_2 +\ddot q_2 \dot q_1)\nn\\
&=& \fft{1}{m+1} (\dot q_1 + \dot q_2)^{m+1} (\ddot q_1 \dot q_2 +\ddot q_2 \dot q_1)
+ \fft{1}{m+1} \big(q_1 q_2 (\dot q_1\dot q_2)^{m+1}\big)\dot{}\,.
\eea
Indeed, as we have shown in the previous section, in the Gaussian coordinates $x^\mu=\{\omega,x^a\}$, we are able to write GB gravity as
\bea
	\sqrt{-g}L_{\text{GB}}&=&
	N\sqrt{h}\delta^{a_1a_2a_3a_4}_{b_1b_2b_3b_4}\bigg(\frac{1}{4}
\bar{R}^{b_1b_2}_{a_1a_2}\bar{R}^{b_3b_4}_{a_3a_4}+\epsilon K^{b_1}_{a_1}K^{b_2}_{a_2}\big(\bar{R}^{b_2b_3}_{a_2a_3}-\frac{\epsilon}{3}K^{b_3}_{a_3} K^{b_4}_{a_4}\big)\bigg)\nonumber\\
	&&
	+\partial_\omega \bigg(-2 \epsilon \sqrt{h} \delta^{a_1a_2a_3}_{b_1b_2b_3} K^{b_1}_{a_1}(\bar{R}^{b_2b_3}_{a_2a_3}-\frac{2 \epsilon}{3}K^{b_2}_{a_2} K^{b_3}_{a_3} )  \bigg),
\eea
in which the bulk term is a first-order Lagrangian \cite{Tel,Pap,Fel}. Furthermore it can be written in terms of a classical holographic relation as Eq.~(\ref{hol}) where the bulk term is a first order Lagrangian in terms of $w$ derivatives. The bulk term in this ADM formalism is no longer the same as the bulk term in the foliation-independent approach.

\subsection{General $k$'th Lovelock gravity}

Following the same approach of GB gravity, we can find that Lovelock gravity is in general classical holographic in the foliation-independent formalism, namely
\begin{align}
	\sqrt{-g}L_{\rm Lovelock}^{(k)}(\Gamma, \partial \Gamma)=\sqrt{-g}\mathit{L}_{\text{ bulk}}^{(k)}(\Gamma, \partial \Gamma)
	-\partial_\mu \Big(\delta ^\beta_\rho \frac{\partial(\sqrt{-g} L^{(k)}_{\text{ bulk}})}{\partial \Gamma^\beta {}_{\mu \rho}}+	\Gamma^\beta {}_{\alpha \rho} \frac{\partial(\sqrt{-g} L^{(k)}_{\rm bulk})}{\partial (\partial_\mu \Gamma^\beta{} _{\alpha \rho})}\Big).
	\label{h1}
\end{align}
To be specific, it is instructive to define
\be
L^{(k)}_0=\delta^{\mu_1\cdots \mu_{2k}}_{\nu_1\cdots \nu_{2k}}\,
	(\Gamma^2)^{\nu_1\nu_2}_{\mu_1\mu_2} (\Gamma^2)^{\nu_3\nu_4}_{\mu_3\mu_4} \cdots
	(\Gamma^2)^{\nu_{2k-1}\, \nu_{2k}}_{\mu_{2k-1}\, \mu_{2k}}\,,
\ee
and define $L^{(k)}_i$ as $L^{(k)}_0$ with ``$i$'' $(\Gamma^2)$ factors replaced by the $(D\Gamma^2)$ terms of the same indices; therefore, we have $i=0,1,\ldots, k$. It is then clear that we have
\be
L_{\rm Lovelock}^{(k)}=\sum_{i}^k \binom{k}{i} L^{(k)}_i\,.
\ee
For two low-lying examples, we have
\be
R = L^{(1)}_0 + L^{(1)}_1\,,\qquad L_{\rm GB}=L^{(2)}_0 + 2 L^{(2)}_1 + L^{(2)}_2\,.
\ee
We find that the corresponding bulk Lagrangian is given by
\be
L_{\rm bulk}^{(k)}=-\sum_{i}^{k-1} \binom{k-1}{i} L^{(k)}_i\,.\label{genbulk}
\ee
The minus sign is intriguing and for $k\ge 2$, it is not the same as the bulk term in the ADM formalism given in Eq.~\eqref{hol}.

It is now worth comparing our results to those in literature. For a general Lovelock Lagrangian, without calculating the bulk term, the following surface Lagrangian in terms of metric has been proposed \cite{pad1,pad2}
\begin{equation}\label{pady1}
		\partial_\mu J^{\mu (k)}=-\frac{1}{(D/2)-2k} \partial_\mu \bigg(g_{\nu \rho} \frac{\partial (\sqrt{-g}L^{(k)}_{\rm bulk})}{\partial( \partial_\mu g_{\nu \rho}) }  + \partial_{\delta} g_{\nu \rho} \frac{\partial (\sqrt{-g}L^{(k)}_{\rm bulk})}{\partial( \partial_\mu \partial_{\delta} g_{\nu \rho})} \bigg).
	\end{equation}
To prove above equation, the homogeneity of the bulk expression was assumed, with the following relations for the bulk term
	\begin{align}
		&g_{\mu \nu} \frac{\partial (\sqrt{-g} L^{(k)}_{\rm bulk})}{\partial g_{\mu \nu}}= ((D/2)-n_0) \sqrt{-g} L^{(k)}_{\rm bulk}, \hspace{0.8cm} \partial\rho g_{\mu \nu} \frac{\partial (\sqrt{-g} L^{(k)}_{\rm bulk})}{\partial( \partial\rho g_{\mu \nu})}=n_1 \sqrt{-g} L^{(k)}_{\rm bulk},
		\nonumber \\
		& \partial_\delta \partial_\rho g_{\mu \nu} \frac{\partial (\sqrt{-g} L^{(k)}_{\rm bulk})}{\partial( \partial\delta \partial_\rho g_{\mu \nu})}=n_2 \sqrt{-g} L^{(k)}_{\rm bulk},
		\label{homo}
	\end{align}
	in which $n_i$'s are number of the factors of  $g_{\mu \nu}$, $\partial_\rho g_{\mu \nu}$ and $\partial_\delta \partial_\rho g_{\mu \nu}$ in any given term of the bulk. However, considering the bulk term presented in Eq.~\eqref{genbulk}, it is obvious that there is no homogeneity in $\partial_\rho g_{\mu \nu}$ and $\partial_\delta \partial_\rho g_{\mu \nu}$, so we are not allowed to write the identities in Eq.~(\ref{homo}) for the bulk term. In the next subsection, we give explicitly examples to illustrate that Eq.~\eqref{pady1} is incorrect.

\subsection{FLRW model as a concrete example}

We now consider the FLRW model as a concrete example to illustrate the difference between the classical holographic relations, derived from the ADM formalism and the foliation-independent formalism.  We consider the cosmological metric in general $D$ dimensions
\be
ds^2 = - dt^2 + a(t) dx^i dx^i\,.
\ee
Note that here we use $a(t)$ rather than $a(t)^2$ as the scaling factor, so that the metric is $g_{ij}=a \delta_{ij}$. Note that we can treat this metric as special case of the general foliation-independent approach, or alternatively as a special case of the ADM decomposition where the special coordinate $w=t$. This allows us to compare the bulk Lagrangian in two approaches.
For this cosmological ansatz, we have
\bea
&&\sqrt{-g} L^{(k)}_0 = \fft{(D-1)!}{(D-2k-1)!} a^{D-2k-1} \dot a^{2k}\,,\qquad
L^{(k)}_1 = 2\fft{(D-1)!}{(D-2k)!} a^{D-2k} \dot a^{2(k-1)} \ddot a\,,\nn\\
&&\sqrt{-g} L^{(k)}_i = 0\,,\qquad i=2,3,\ldots, k\,.
\eea
The general Lovelock and its bulk Lagrangian are therefore given by
\be
\sqrt{-g} L^{(k)}_{\rm Lovelock} = \sqrt{-g} \big(L^{(k)}_0 + k L^{(k)}_1\big)\,,\qquad
\sqrt{-g} L^{(k)}_{\rm bulk} = -\sqrt{-g} \big(L^{(k)}_0 + (k-1) L^{(k)}_1\big)\,.
\ee
It is straightforward to verify that
\be
\sqrt{-g} L^{(k)}_{\rm Lovelock}=\sqrt{-g} L^{(k)}_{\rm bulk}  + \partial_t J^t\,,
\ee
where
\be
J^t = 2\fft{(D-1)!}{(D-2k)!} a^{D-2k} \dot a^{2k-1}\,.
\ee
The fact that the difference between the full Lagrangian and our bulk term gives a total derivative confirms our bulk Lagrangian formula Eq.~\eqref{genbulk}, even though the FLRW metric is somewhat too simple to be a proof. For the classical holographic relation, we propose the ansatz on the metric function
\be
J^t = c_1 a \fft{\partial (\sqrt{-g} L_{\rm bulk}^{(k)})}{\partial \dot a} +
 c_2 \dot a \fft{\partial (\sqrt{-g} L_{\rm bulk}^{(k)})}{\partial \ddot a},
\ee
where $(c_1,c_2)$ are numerical coefficients that are to be determined.  We find the above relation holds provided that
\be
4c_1 (k-1)^2 a \ddot a + \Big(2(k-1) c_2 + (D+2-4k)k c_1 + 2\Big) \dot a^2=0\,.
\ee
Thus we have
\bea
k=1:&&\qquad c_1=-\fft{2}{D-2}\,,\qquad \hbox{$c_2$ is irrelevant;}\nn\\
k\ge 2:&&\qquad c_1=0\,,\qquad c_2 = - \fft{1}{k-1}\,.
\eea
Thus for the cosmological model, we have
\bea
&&\sqrt{-g} R = \sqrt{-g} L_{\rm bulk}^{(1)} - \fft{2}{D-2} \fft{d}{dt}\Big(a \fft{\partial (\sqrt{-g} L_{\rm bulk}^{(1)})}{\partial \dot a}\Big) \,,\nn\\
&&\sqrt{-g} L_{\rm Lovelock}^{(k)} =\sqrt{-g} L_{\rm bulk}^{(k)} - \fft{1}{k-1} \fft{d}{dt}\Big(\dot a\fft{\partial (\sqrt{-g} L_{\rm bulk}^{(k)})}{\partial \ddot a}\Big)\,,\qquad k\ge 2\,.
\eea
To compare this classical holographic relation to that in the ADM formalism, we note that
\be
\sqrt{-g} L^{(k)}_1 = - \fft{2}{2k-1} L^{(k)}_0 - \fft{2}{k (2k-1)(D-2k)}\,
\fft{d}{dt}\Big(a \fft{\partial (\sqrt{-g} L^{(k)}_0)}{\partial \dot a}\Big)\,.
\ee
This leads to the classical holographic relation in the ADM formalism, namely
\be
\sqrt{-g} L_{\rm Lovelock}^{(k)} = \sqrt{-g} L^{(k)}_{\rm ADM} -\fft2{D-2k}\fft{d}{dt}
\Big(a \fft{\partial (\sqrt{-g} L^{(k)}_{\rm ADM})}{\partial \dot a}\Big)\,,
\ee
with $L^{(k)}_{\rm ADM} = - \fft{1}{2k-1}L^{(k)}_0$. This example explicitly shows the proposed surface term in (\ref{pady1}) is not correct. Note that for the ADM approach, the form of the classical holographic relation is ``continuous'' as $k$ running from 1 to higher values. However, the relation is discontinuous in the foliation-independent approach when we use $(a, \dot a, \ddot a)$ as variables. It is also worth emphasizing that for $k=1$, we have $L_{\rm bulk} = L_{\rm ADM}$. However, for $k\ge 2$, $L_{\rm bulk}$ in the foliation-independent approach involves $\ddot a$, as well as $(a,\dot a)$; on the other hand, $L_{\rm ADM}$ is first order involving only $(a,\dot a)$.

We further consider the most general cosmological ansatz in $D=5$: $ds_5^2=g_{\mu\nu} (t) dx^\mu dx^\nu$, with $x^\mu = (t, x_1, x_2, x_3, x_4)$. We follow Eq.~\eqref{pady1} and consider the ansatz for GB gravity ($k=2$)
\begin{equation}\label{pady2}
		\partial_t J^{t (2)}=-\frac{1}{(5/2)-4} \partial_t \bigg(c_1 g_{\mu\nu} \frac{\partial (\sqrt{-g}L^{(2)}_{\rm bulk})}{\partial( \partial_t g_{\mu\nu}) }  + c_2\partial_{t} g_{\mu\nu} \frac{\partial (\sqrt{-g}L^{(2)}_{\rm bulk})}{\partial( \partial_t \partial_{t} g_{\mu\nu})} \bigg),
	\end{equation}
with two arbitrary numerical coefficients $(c_1,c_2)$. We use computer to obtain explicit $L_{\rm bulk}$ and we find that there is no solution for $(c_1,c_2)$.

\section{Conclusion}

In this paper, we obtained the classical holographic relations for the general Lovelock Lagrangian following two approaches: one based on the ADM formalism and the other with theO surface term $\partial_\mu J^\mu$. In the latter foliation-independent approach, the bulk Lagrangian depends not only $\Gamma$, but also $\partial \Gamma$ for $k\ge 2$; therefore, it is not first order, or degenerate. Our classical holographic relation Eq.~\eqref{h1} differs from Eq.~\eqref{pady1} that was in literature and we used explicit examples to demonstrate that Eq.~\eqref{pady1} is incorrect.

The non-degenerate issue can be resolved by using the ADM decomposition $x^\mu=(w,x^a)$ where the foliation coordinate $w$ is treated as special. Using the results of \cite{Tel,Pap,Mayers} for the first-order Lagrangians and boundary terms, we obtained the classical holographic-degenerate relation \eqref{hol} for the Lovelock Lagrangian. By first order, we mean that the bulk Lagrangian has no more than one derivatives of $w$; its two-derivatives terms are all associated with the derivatives of $x^a$, entering the Lagrangian through the Riemann tensor of the sub-manifold $x^a$. The bulk Lagrangians from both the ADM and foliation-independent approaches turn out to be same for Einstein gravity, but they are not for general Lovelock gravities.

The resulting classical holographic-degenerate relation in the ADM formalism enabled us to compute the appropriate Myers terms consistent with the Dirichlet or Neumann BCs (related by a factor dependent on the space-time dimension). We found this interesting result that in $D=2(k+1)$, the Myers term under Neumann BC vanishes for the $k$'th-order Lovelock gravity. As a contrast, in a general foliation independent approach, {\it i.e.} the foliation-independent approach discussed earlier, our result is different and the Lovelock Lagrangian is not degenerate, except for $k=1$, since bulk term contains second-order derivatives. We used the FLRW cosmological metric in general dimensions to illustrate the differences between the ADM formalism and the foliation-independent approach.

For generalization, we considered $f (L_{\rm Lovelock}^{(k)})$-gravity and figured out its Lagrangian is not classical holographic. Also it can not be expressed as the sum of the first-order and the total derivative terms. So $f (L_{\rm Lovelock}^{(k)})$ Lagrangian is not degenerate. Following the Ostrogradsky approach \cite{Os} and writing $f(L_{\rm Lovelock}^{(k)})$-gravity action in the frame work of Brans-Dicke formalism \cite{Cap, Sot, Far} gave us the degenerate Lagrangian which we used to develop the issue of BCs and the corresponding Myers terms. Here we followed ADM formalism to find the appropriate Myers terms in $f(L_{\rm Lovelock}^{(k)})$-gravity, required to make the variation principle well defined. We have introduced Neumann BC and two types of mixed BCs in addition to the Dirichlet BC. The remarkable result is that under a type of mixed BC we do not need to add any Myers term to the action of $f (L_{\rm Lovelock}^{(k)})$-gravity in all $D$ dimensions.

Moreover, we investigated physical quantities such as free energy, energy and black hole entropy, in the framework of Euclidean semiclassical approximation method \cite{Hint,HKH2} {\it via} different BCs in Lovelock gravity and $f (L_{\rm Lovelock}^{(k)})$-gravity. In this method the black hole entropy may be obtained for different kinds of BCs and Myers terms. However, we showed that the main term, which was responsible to give the difference of the numerical value of the action with the background solution, was the same for all cases, regardless of the particular kind of BCs. In fact, decomposing the space-times boundary into two space-like $\Sigma_t$ and one time-like $\mathcal{B}$ hyper-surfaces, one can see the integral over $\mathcal{B}$ has no role in making the action principle well-defined for asymptotically flat metrics \cite{HKH1,HKH2}. However, to calculate the black hole entropy, the integral over $\mathcal{B}$ is exactly the term which gives nontrivial contributions for different BCs. Taking this point into account, we showed in Lovelock gravity, under Dirichlet and Neumann BCs and also in $f (L_{\rm Lovelock}^{(k)})$-gravity under Dirichlet, Neumann and two types of mixed BCs, free energy, energy and the entropy do not change. In our approach the actions do not change for all stationary backgrounds, but they do depend on the BCs in time-dependent space-times; its physical implication in cosmology needs to be investigated.

\section*{Acknowledgement}

We are grateful to Zhan-Feng Mai for useful discussions. This work is supported in part by the National Natural Science Foundation of China (NSFC) grants No.~11875200 and No.~11935009.

\end{document}